# Dynamic Consequences of Optical Spin-Orbit Interaction

Sergey Sukhov, Veerachart Kajorndejnukul, and Aristide Dogariu

*CREOL, The College of Optics and Photonics University of Central Florida, 4000 Central Florida Boulevard*

*Orlando, Florida 32816, USA*

When circularly polarized wave scatters off a sphere, the scattered field forms a vortex with spiraling energy flow. This is due to the transformation of spin angular momentum into orbital one. Here we demonstrate that during this scattering an anomalous force can be induced that acts in a direction perpendicular to the propagation of incident wave. The appearance of this lateral force is made possible by the presence of an interface in the vicinity of scattering object. Besides radiation pressure and tractor-beam pulling forces, this lateral force is another type of non-conservative force that can be produced with unstructured light beams.

## I. INTRODUCTION

Upon interaction with matter, there is an exchange between the spin and the orbital parts of the momentum carried by an optical wave. This optical spin-orbit interaction is responsible for a number of wave polarization effects [1,2,3,4,5]. Moreover, the conservation of total momentum also involves momentum transfer to matter. When analyzing this conservation law, the symmetry of the field is a critical component. For instance, when the azimuthal symmetry of the field is preserved around an axis, the projection of the total angular momentum along that axis is conserved according to Noether's theorem. As a result, the mechanical action on matter is along this axis of symmetry. When the rotational symmetry is broken as a result of interaction, the direction of the mechanical action is affected in order to obey the conservation of canonical momentum.

The field mirror symmetry can be perturbed even when interacting with rotationally symmetric objects; a vortex field emerges centered at the location of a spherical scatterer when a circularly polarized plane wave is incident on it [6]. This vortex structure arises because of a partial transformation of spin to orbital angular momentum. Previously, we have demonstrated that, because of this spiraling around a sphere, the power flow experiences a shift analogous to the spin-Hall effect of light [7]. In [7] the field asymmetry during spin-orbit interaction (SOI) was tested by directly measuring the light to the right and to the left of a scattering particle. A variation of this experiment was later proposed [8,9] by replacing the measurement fiber with different waveguide structures or interfaces. In these situations the rotational symmetry was broken by the presence of nearby waveguides.

The field symmetry can also be broken when more scattering objects interact optically. For example, we have shown that a test object placed in the vicinity of scattering sphere should experience the radiation pressure from the curved power flow causing both objects to be affected by spin and orbital torques [10]. However, as we said above, the asymmetry of light scattering during SOI can be affected by nearby interfaces and this could lead to transversal forces acting on the object. To quantitatively understand this, let us consider a spherical particle located at the interface of two media with different refractive indices and illuminated with a circularly polarized wave. Figure 1 presents numerical calculations of this situation. As can be seen, the spiraling of energy flow breaks the mirror symmetry of light scattering. Because of optical spin Hall in spherical geometry [7], light to left of the particle propagates mostly in the upper medium while the opposite happens to the right side of the particle. The presence of the interface further breaks the vortex central symmetry. As the magnitude of wave's momentum is determined by the properties of medium it propagates through [[11]], this scattering asymmetry unbalances the transversal linear momentum. Consequently, a side force perpendicular to the original direction of propagation should act on the particle with a magnitude determined by its scattering phase-function [12]. In the present Letter, we further elaborate this idea analytically and confirm it by exact numerical calculations.

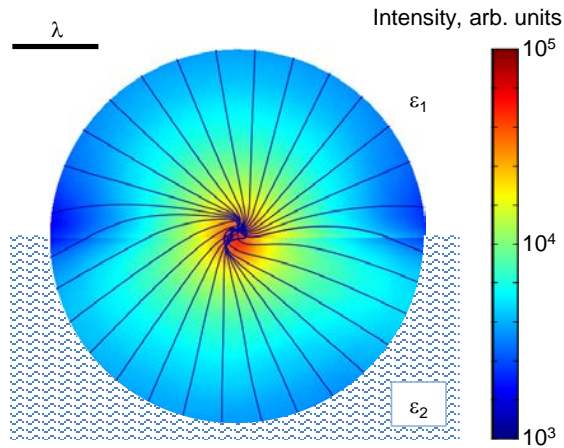

**Figure 1.** Intensity distribution (color) and power flow (lines) for a dipole located at the interface of two media with dielectric permittivities $\varepsilon_1$ and $\varepsilon_2$. Calculations were performed in Comsol Multiphysics for a 100nm TiO$_2$ particle at water-air interface. The particle is illuminated by circularly polarized plane wave ($\lambda$=532nm) incident at 55° with respect to the normal.

## II. SINGLE DIPOLE NEAR A DIELECTRIC INTERFACE

If a dipolar particle is located entirely inside one of the two adjacent media, the problem can be solved analytically. We consider a point dipole located a distance $z_0$ away from the interface between two semi-infinite transparent media 1 and 2 (Figure 2). Let us assume that the dipole is placed in medium 2 and that a circularly polarized plane wave of amplitude $E_{0I}$ is incident from medium 1 (Figure 2). We want to determine the optical forces acting on the dipole. In particular, we are interested in the force $F_\perp$ acting in a direction perpendicular to the plane of incidence $xz$. Obviously, $F_\perp$ is zero for normal incidence $\theta_I = 0°$ because of the field azimuthal symmetry. The force also becomes zero for slanted angles of incidence $\theta_I \to 90°$ as no light reaches the dipole inside the medium 2. However, for intermediate angles, $F_\perp$ has finite values as will be shown in the following.

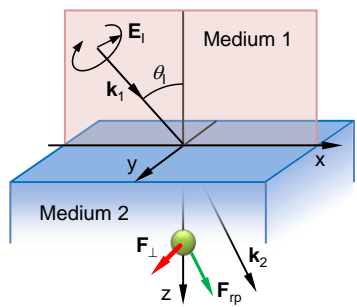

**Figure 2.** Geometry of the problem. A dipole size particle in medium 2 is illuminated by a refracted circularly polarized wave incident from medium 1. Besides the usual radiation pressure $\mathbf{F}_{rp}$, the particle experience a side force $\mathbf{F}_\perp$ determined by a presence of the interface.

The force acting on the dipole can be found in a usual way [13,14]

$$F_\perp = \frac{1}{2}\text{Re}(\mathbf{d}\,\partial_y \mathbf{E}^*), \qquad (1)$$

where $\mathbf{d} = \alpha \mathbf{E}$ is the induced dipole moment with $\alpha$ being the polarizability, $\mathbf{E} = \mathbf{E}_T + \mathbf{E}_d$ is the total field acting on a dipole, $\mathbf{E}_T$ is the field strengths of the transmitted wave, $\mathbf{E}_d$ is the dipole field scattered from the interface back to the location of the dipole. The symbol $\partial_y$ denotes partial derivative with respect to transversal coordinate $y$. As there is no dependence on a transversal coordinate for the transmitted wave, the derivative in Eq. (1) is $\partial_y \mathbf{E}_T^* = \partial_y \mathbf{E}_d^*$. This derivative can be readily calculated knowing the Green's function $\mathbf{G}$ of a dipole radiation near the surface [15]. With this, the final expression for the force (1) can be written as

$$F_\perp = \frac{k^2}{\varepsilon_0}\,\text{Im}(d_y d_z^*)\,\text{Im}(\partial_y G_{yz}^0), \quad \partial_y G_{yz}^0 = \frac{1}{8\pi k_2^2}\int_0^\infty r_p(k_\rho)k_\rho^3 \exp(2ik_{2z}z_0)dk_\rho. \qquad (2)$$

Here $G_{yz}^0 = G_{yz}(\mathbf{r}_0,\mathbf{r}_0)$ is the $yz$ component of the Green's function tensor [15], $r_p$ is the amplitude reflection coefficient for p-polarized wave, $k_2$ is the wave number in medium 2, $k_{2z} = \sqrt{k_2^2 - k_\rho^2}$, $k$ is the wavenumber in vacuum. We must stress that as there are no variations of intensity along the surface, Eq.(2) describes purely nonconservative force. Furthermore, one can see that the force is determined by two factors: the first one is specified by configuration of the field in the vicinity of interface, the second one is entirely determined by the material properties of media.

For the force in Eq.(2) to be nonzero, there should be a phase difference between the y- and z-components of dipole momentum. This practically means that this transversal force exists only for elliptically polarized light and disappears in the case of linearly polarized wave. The magnitude of the dipole moment in Eq.(2) should be found self-consistently accounting for the influence of both the refracted wave and field $\mathbf{E}_d$. However, for dielectric media the influence of the latter is insignificant and $\text{Im}(d_y d_z^*)$ can be written as

$$\text{Im}(d_y d_z^*) = \pm \tfrac{1}{2}|\alpha|^2\,E_{0I}^2 t_s(\theta_I)t_p(\theta_I)\sqrt{\varepsilon_{12}}\,\sin\theta_I. \qquad (3)$$

The plus sign in Eq.(3) corresponds to the left circular polarization, the minus sign indicates the right one; $t_s$, $t_p$ are the amplitude transmission coefficients for s- and p-polarized waves, correspondingly, $\alpha$ is a dipole polarizability, and $\varepsilon_{12} = \varepsilon_1/\varepsilon_2$ is the relative dielectric permittivity of two media. One can see from Eq.(3), that force $F_\perp$ has the same dependence on polarizability ($\alpha^2$) as radiation pressure.

For the case of dipole placed on a surface ($z_0 \to 0$) the integral in Eq.(2) can be calculated directly and one obtains an explicit expression for the transversal force. For example, in the case of dipole located in more optically dense medium ($\varepsilon_{12} < 1$) one can get:

$$\text{Im}\left(\partial_y G^0_{yz}\big|_{z_0=0}\right) = k_2^2 \frac{\varepsilon_{12}}{64} \frac{1-\varepsilon_{12}}{1+\varepsilon_{12}} \left(1 + \frac{4\varepsilon_{12}}{(1+\varepsilon_{12})^2}\right). \quad (4)$$

It is interesting to notice that even though the electrostatic approximation can be used to calculate the dipole moments for $k_2 z_0 \ll 1$, it is unsuitable for calculation of our transversal nonconservative force. Electrostatic approximation does not include retardation effects (field phases) and, thus, can account only for conservative forces. This demonstrates that limitations of electrostatic approximation are even more severe than previously believed.

For typical dielectric permittivities ($\varepsilon_{12} \in [0.1, 0.6]$), Eq.(4) takes values $\sim 0.01 k_2^2$. Combining this with Eqs.(2) and (3), one can conclude that transversal force is expected to be about two orders of magnitude smaller than the radiation pressure.

For a dipole far from interface ($k_2 z_0 \gg 1$), the integral in Eq.(2) can be evaluated asymptotically (see Supplementary information for details) to obtain

$$F_\perp(z_0 \to \infty) \sim \frac{k_2^4}{4\pi\varepsilon_2\varepsilon_0} \text{Im}(d_y d_z^*) \frac{\sin(2k_2 z_0)}{(2k_2 z_0)^2} \frac{n_{12}-1}{n_{12}+1}. \quad (5)$$

The result in Eq.(5) suggests another interpretation of the transversal force for particles located far from interface. The vortex spherical wave created by a dipole is got reflected from the interface and causes back-influence on the dipole. The force can be treated as appearing from the phase of the vortex. The same phase, for example, makes particles orbiting in vortex beams [16,17]. In our case the particle creates vortex and drags it with itself creating self-propelling linear motion. This interpretation becomes similar to the one suggested for chiral particles at an interface [18]. However, our demonstration indicates that this phenomenon is more general and does not necessarily rely on exotic material properties.

Furthermore, the oscillating behavior of $F_\perp$ with distance $z_0$ resembles the one occurring in optical binding. Previously, we have suggested that transversal force will appear for a system of optically bound particles illuminated with circular polarization [10]. However, the distance-dependent force was proportional to the inverse third power of the interparticle distance [10] and decayed faster than the force in Eq.(5).

In a similar way one can estimate the transversal force for a particle located in the same medium as incident wave (see Supplementary materials). However, in this case one should be careful in interpreting the nature of this transversal force. The force in Eq.(2) should be distinguished from the action of a spin momentum that can appears during the interference of incident and reflected waves [19,20,21]. Notably, the spin momentum does not affect dipolar particles and it can be detected only in the presence of multipoles. In fact, it is this type of transversal force that was measured in our experiment on Aharonov-Bohm effect in optical setting for 2μm particles [12]. As opposed to the force emerging from the spin momentum, the transversal force described in this Letter would be preserved even if the reflected wave is eliminated by, for example, an antireflection coating at the interface.

### III. NUMERICAL ESTIMATIONS FOR THE MAGNITUDE OF TRANSVERSAL FORCE

One can see from Eq.(5) that the transversal force is long range and thus can affect dipoles located far from the surface. This suggests that such transversal forces should also appear for objects with dimensions larger than a wavelength. To gain some insight about the magnitude of this transversal force for larger particles, we performed numerical simulations (Comsol Multiphysics) for a structured particle similar to those used in our experiment [22]. Figure 3a shows the transversal force on a 1.5µm particle as a function of the angle of incidence. For irradiances of 1mW/µm$^2$ the force reaches values of tens of fN that can be detected experimentally. Figure 3b shows the vertical cross-section perpendicular to the plane of incidence for the field distribution around the particle. One can see the asymmetric power flow around the particle. Also one can see transversal spin momentum flow in an interference pattern of incident and reflected waves. The maximum flow occurs away from the surface and is expected not to contribute much to the transversal force of interest.

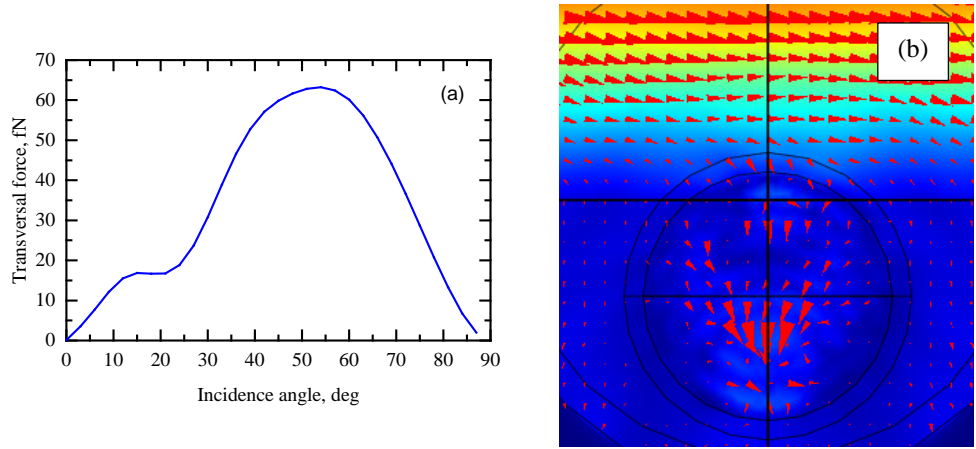

**Figure 3.** (a) Transversal force as a function of the angle of incidence; (b) Intensity (color) and power flow (arrows) in a transversal vertical cross-section perpendicular to the plane of incidence for incidence angle $\theta_I = 55°$. The 1.5µm particle consists of 100nm thick TiO$_2$ coating (refractive index 2.67) and silica core (refractive index 1.46) and located at water-air interface. The contact angle between TiO$_2$ and water was set to be $48°$.

## IV. CONCLUSIONS

We presented evidence for a novel manifestation of optically-induced non-conservative force. This new force completes the set of methods available for arbitrary manipulation of colloidal particles on a surface of liquids. The forward movement can be provided by radiation pressure, the backward movement can be achieved by a momentum enhancement effect [11]. Here we have shown that a side motion can be provided by the transformation of spin angular momentum of incident photons into orbital angular momentum of scattered ones.

# Supplementary Materials

## EVALUATION OF THE INTEGRALS ENTERING THE EXPRESSION FOR DERIVATIVE OF A GREEN'S FUNCTION

To gain some insight about the magnitude and distance dependence of transversal force in Eq.(2) of the main text, one should evaluate the integral entering this expression. For convenience, we switch to dimensionless parameters $s = k_\rho / k_2$, $\xi = k_2 z_0$. The integral in Eq.(2) of the main text can be rewritten as

$$\partial_y G_{yz}^0 = \frac{k_2^2}{8\pi} \int_0^\infty r_p(s) s^3 \exp(2i\xi\sqrt{1-s^2}) ds \qquad (S.1)$$

with amplitude reflection coefficient defined by the following expression:

$$r_p(s) = \frac{\sqrt{\varepsilon_{12} - s^2} - \varepsilon_{12}\sqrt{1-s^2}}{\sqrt{\varepsilon_{12} - s^2} + \varepsilon_{12}\sqrt{1-s^2}}.$$

Here $\varepsilon_{12} = \varepsilon_1 / \varepsilon_2$ is the relative dielectric permittivity between two media with index "2" corresponding to the medium containing dipole, $k_2$ is the wavenumber in a medium of dipole's location. For evaluation of integral in Eq.(S.1) we need to consider separate cases of $\varepsilon_{12} > 0$ and $\varepsilon_{12} < 0$.

   a) *Dipole in optically less dense medium ($\varepsilon_{12} > 0$)*

For the *case of a dipole located close to the interface* between two media ($\xi \to 0$), (S.1) can be evaluated analytically. As we are interested only in imaginary part of the integral (S.1), the integration can be limited only to interval $[1, \sqrt{\varepsilon_{12}}]$ where reflection coefficient $r_p(s)$ is complex-valued:

$$\mathrm{Im}\left(\partial_y G_{yz}^0\right) = -\frac{k_2^2}{8}\varepsilon_{12}\left[\frac{1}{8}\left(\frac{\varepsilon_{12}-1}{\varepsilon_{12}+1}\right)^3 - \frac{3\varepsilon_{12}^2 - \sqrt{\varepsilon_{12}^3(2+5\varepsilon_{12}+2\varepsilon_{12}^2)}}{(\varepsilon_{12}-1)(\varepsilon_{12}+1)^3}\right], \quad \xi \to 0. \quad (S.2)$$

We would like to stress again that even though $k_1 z_0 \ll 1$, the transversal force cannot be obtained from electrostatic approximation.

The magnitude of $\mathrm{Im}\left(\partial_y G_{yz}^0\right)$ monotonously decreases with increase of $\varepsilon_{12}$ reaching $\mathrm{Im}\left(\partial_y G_{yz}^0\right) \approx -k_2^2/10$ for $\varepsilon_{12} \approx 10$. Thus, for particles in less dense medium, transversal force is expected to be an order of magnitude smaller than radiation pressure.

For the *case of large $\xi$* the integration can be limited only to the interval $[0, \sqrt{\varepsilon_{12}}]$ where at least one of the factors under the integral is complex-valued. Reflection coefficient $r_p(s)$ is purely real in the interval [0, 1]. In this interval the integral in Eq.(S.1) can be evaluated by the method of stationary phase [1,2] with stationary point $s_s = 0$ coinciding with one of the bounds of the integral. The contribution from non-stationary end-point $s = 1$ can be found as a limit $\lim_{\delta \to o} \int_0^{1-\delta} (...)ds$ to avoid divergences.

The region of integration $[1, \sqrt{\varepsilon_{12}}]$ corresponds to evanescent waves of a dipole that are converted into propagating ones after passing through interface. In this interval otherwise oscillating factor $\exp(2i\xi\sqrt{1-s^2})$ becomes exponentially decaying. The replacement of variables $t^2 = s^2 - 1$ converts this integral into Laplace one that can be evaluated asymptotically [3].

Limiting ourselves to the terms up to $1/\xi^3$, we can get the following expression for the derivative of Green's function:

$$\text{Im}(\partial_y G_{yz}^0) \sim -\frac{k_2^2}{8\pi}\left[\frac{n_{12}-1}{n_{12}+1}\left(\frac{\sin 2\xi}{2\xi^2} + \frac{\cos 2\xi}{\xi^3}\left(\frac{3}{4} + \frac{1}{n_{12}}\right)\right) + \frac{\varepsilon_{12}}{2\xi^3\sqrt{\varepsilon_{12}-1}}\right], \quad \xi \to \infty, \quad (S.3)$$

where $n_{12} = \sqrt{\varepsilon_{12}}$ is the relative refractive index. Interestingly, Eq.(S.3) includes both retarded (depending on a phase $2\xi$) and nonretarded non-oscillating $1/\xi^3$ terms.

b) *Dipole in optically more dense medium ($\varepsilon_{12} < 0$)*

The expression for the *case of a dipole close to the interface* ($\xi \to 0$) was already shown in the main text:

$$\text{Im}(\partial_y G_{yz}^0) = \frac{k_2^2}{64}\varepsilon_{12}\frac{1-\varepsilon_{12}}{1+\varepsilon_{12}}\left(1 + \frac{4\varepsilon_{12}}{(1+\varepsilon_{12})^2}\right), \quad \xi \to 0.$$

This expression can be obtained analytically by direct integration of Eq.(S.1) over the interval $[\sqrt{\varepsilon_{12}}, 1]$ where reflection coefficient $r_p(s)$ is complex-valued. The typical magnitude of this expression is $\text{Im}(\partial_y G_{yz}^0) \approx k_2^2/100$ that makes the case of the particle located in optically dense medium less favorable in terms of the magnitude of transversal force than the case of less dense medium.

For the *case of large $\xi$* the integration can be limited to the interval $[0, 1]$ with a special attention given to the point $\sqrt{\varepsilon_{12}}$ where reflection coefficient transforms from purely real to complex-valued. It is convenient to rewrite the reflection coefficient as

$$r_p(s) = 1 - \frac{2\varepsilon_{12}\sqrt{1-s^2}}{\sqrt{\varepsilon_{12}-s^2} + \varepsilon_{12}\sqrt{1-s^2}}. \quad (S.4)$$

Integration with the first term in Eq.(S.4) can be performed analytically. The obtained expression by itself surprisingly well reproduces the exact behavior of integral (S.1) for the case of high dielectric contrast ($\varepsilon_{12} \ll 0$)

$$\mathrm{Im}\bigl(\partial_y G_{yz}^0\bigr) \approx \frac{k_2^2}{8\pi} \frac{(3-4\xi^2)\sin(2\xi) - 6\xi\cos(2\xi)}{8\xi^4}, \quad \varepsilon_{12} \ll 0. \quad (S.5)$$

This expression works well for arbitrary particle-interface separations $\xi$. The asymptotic evaluation of the integral (S.1) with the second term of (S.4) can be again performed by the method of stationary phase [1]. When performing calculations, one might need to use substitution of the integration variable $s \to \sqrt{t^2 + \varepsilon_{12}}$ or $s \to \sqrt{\varepsilon_{12} - t^2}$ (depending on the domain of integration) to avoid divergences in the end points. Performing asymptotic evaluation with accuracy up to $1/\xi^2$ terms, one can get

$$-2\varepsilon_{12} \int_0^\infty \frac{\sqrt{1-s^2}\, s^3 \exp(2i\xi\sqrt{1-s^2})}{\sqrt{\varepsilon_{12}-s^2} + \varepsilon_{12}\sqrt{1-s^2}} ds \sim \frac{n_{12}}{1+n_{12}} \frac{\sin 2\xi}{\xi^2}. \quad (S.6)$$

Combining Eq.(S.5) and Eq.(S.6) and leaving only terms proportional to $1/\xi^2$, one gets the final asymptotic evaluation of the derivative of the Green's function:

$$\mathrm{Im}\bigl(\partial_y G_{yz}^0\bigr) \sim \frac{k_2^2}{8\pi} \frac{\sin(2k_2 z_0)}{2(k_2 z_0)^2} \frac{n_{12}-1}{n_{12}+1}, \quad \xi \to \infty.$$

Comparing this expression to Eq.(S.3), one can see that they coincide up to $1/\xi^2$ terms.

## TRANSVERSAL FORCE IN THE CASE OF A DIPOLE LOCATED IN THE SAME MEDIUM AS INCIDENT WAVE

Here we consider transversal optical force $F_\perp$ acting on a point dipole located in the same medium as incident wave. The dipole is located a distance $z_0$ away from the plane interface between two media. Circularly polarized plane wave is incident onto this interface with incidence angle $\theta_I$. The force acting on a dipole is again determined by Eq.(2) of the main text. However, now dipole moment is determined by not a single, but by a combination of incident and reflected waves $\mathbf{E}_I$ and $\mathbf{E}_R$.

The same conclusions are valid for the transversal force as were for the case of particle beneath the surface. In particular, the force $F_\perp$ is zero for normal incidence because of azimuthal symmetry of the system. The force is also zero for linearly s- and p-polarized light. However, on the contrary to the case of a single refracted wave, the force can be nonzero for the superposition of s- and p-polarized waves as the interference of incident and reflected waves provides phase different between $d_y$ and $d_z$ components of a dipole moment. We will leave this case without much discussion. The force is also nonzero for circularly polarized incident wave with $F_\perp$ changing sign with the change of handedness. It is straightforward to obtain the following expression for $\mathrm{Im}(d_y d_z^*)$:

$$\mathrm{Im}(d_y d_z^*) = \pm\tfrac{1}{2}|\alpha|^2 E_{0I}^2 \sin\theta_I \bigl[1 + r_s(\theta_I)r_p(\theta_I) + (r_s(\theta_I) + r_p(\theta_I))\cos(2k_1 z_0 \cos\theta_I)\bigr]. \quad (S.7)$$

Here again the plus sign corresponds to the left circular polarization, the minus sign indicates the right one. As expected, the interference term $\cos(2k_1 z_0 \cos\theta_I)$ is present in this expression. For dipoles on a surface ($z_0 \to 0$) one can notice that transversal force would be zero not only for normal incidence, but also for very slanted angles of incidence $\theta_I \to 90°$. For these angles the reflectance of surface is almost ideal ($r_s = r_p = -1$) and the factor in square brackets in Eq.(S.7) becomes zero. The physical explanation of this situation is that interference of incident and reflected waves near the surface make light effectively linearly polarized that makes the spin angular momentum of the total field to be zero. However, besides these particular cases, the expression (S.7) is nonzero that makes nonzero the transversal force acting on a particle.

---

[1] J.J Stamnes, Waves in Focal Regions: Propagation, Propagation, Diffraction and Focusing of Light, Sound and Water Waves (Bristol, England: Adam Hilger, 1986) Ch.8.

[2] R.B.Dingle, Asymptotic Expansions: Their Derivation and Interpretation (Academic Press, London&New York, 1973) Ch.5.

[3] A. Erdelyi, Asymptotic expansions (Dover, New York, 1956).